\documentclass[preprint]{revtex4}

\begin{document}

\title
{Two-dimensional phase transition models and $\lambda \phi^4$ 
field theory}
\author{A.~I.~Sokolov} 
\affiliation{Saint Petersburg Electrotechnical University, 
Professor Popov Street 5, Saint Petersburg 197376, Russia}

\begin{abstract}
The overview is given of the results obtained recently in the 
course of renormalization-group (RG) study of two-dimensional 
(2D) models. RG functions of the two-dimensional $n$-vector 
$\lambda \phi^4$ Euclidean field theory are written down up 
to the five-loop terms and perturbative series are resummed by 
the Pad\'e-Borel-Leroy techniques. An account for the five-loop 
term is shown to shift the Wilson fixed point only briefly, 
leaving it outside the segment formed by the results of the 
lattice calculations. This is argued to reflect the influence 
of the non-analytical contribution to the $\beta$-function. 
The evaluation of the critical exponents for $n = 1$, $n = 0$ 
and $n = -1$ in the five-loop approximation and comparison of 
the results with known exact values confirm the conclusion that 
non-analytical contributions are visible in two dimensions. 
The estimates obtained on the base of pseudo-$\epsilon$ expansions 
originating from the 5-loop 2D RG series are also discussed. 

\vspace{0.5cm}
 
 
\end{abstract}
\maketitle

The field-theoretical renormalization-group (RG) approach proved 
to be a powerful tool for calculating the critical exponents and 
other universal quantities of the basic three-dimensional (3D) 
models of phase transitions. Today, many-loop RG expansions for 
$\beta$-functions (six-loop), critical exponents (seven-loop), 
higher-order couplings (four-loop), etc. of the 3D $O(n)$-symmetric, 
cubic, and some other models are known resulting in high-precision 
numerical estimates for experimentally accessible quantities [1-7]. 
The main aim of this paper is to demonstrate how effective (or 
ineffective) is the field-theoretical RG machinery in two dimensions 
where i) the RG series are stronger divergent and ii) singular 
(non-analytic) contributions to RG functions are expected to be 
larger than for 3D systems.  

The Hamiltonian of the model describing the critical behavior of 
various 2D systems reads:
\begin{equation}
H = 
\int d^2x \Biggl[{1 \over 2}( m_0^2 \varphi_{\alpha}^2
 + (\nabla \varphi_{\alpha})^2) 
+ {\lambda \over 24} (\varphi_{\alpha}^2)^2 \Biggr] ,
\label{eq:1} \\
\end{equation}
where $\varphi_{\alpha}$ is a real $n$-vector field, $m_0^2$ is 
proportional to $T - T_c^{(0)}$, $T_c^{(0)}$ being the mean-field 
transition temperature. 

The $\beta$-function and the critical exponents for the model (1) 
are calculated within the massive theory, with the Green function, 
the four-point vertex and the $\phi^2$ insertion being normalized 
in a conventional way:
\begin{eqnarray}
G_R^{-1} (0, m, g_4) = m^2 , \qquad \quad 
{{\partial G_R^{-1} (p, m, g_4)} \over {\partial p^2}}
\bigg\arrowvert_{p^2 = 0} = 1 , \\
\nonumber
\Gamma_R (0, 0, 0, m, g) = m^2 g_4, \qquad \quad  
\Gamma_R^{1,2} (0, 0, m, g_4) = 1. 
\label{eq:2}
\end{eqnarray}

Since the four-loop RG expansions at $n = 1$ have been obtained 
many years ago [1], we are in a position to find corresponding series 
for arbitrary $n$ and to calculate the five-loop terms. The results 
of our calculations are as follows [8]:    
\begin{eqnarray}
{\beta (g) \over 2} = - g + g^2 - {g^3 \over (n + 8)^2} 
\biggl( 10.33501055 n + 47.67505273 \biggr) \ \qquad \qquad \qquad \qquad
\nonumber \\
+ {g^4 \over (n + 8)^3} \biggl( 5.000275928 n^2 
+ 149.1518586 n + 524.3766023 \biggr) \ \qquad \qquad \qquad \qquad
\nonumber \\
- {g^5 \over (n + 8)^4} \biggl( 0.0888429 n^3 + 179.69759 n^2   
+ 2611.1548 n + 7591.1087 \biggr) \ \ \qquad \qquad 
\nonumber \\
+ {g^6 \over (n + 8)^5} 
\biggl( -0.00408 n^4 + 80.3096 n^3   
+ 5253.56 n^2 + 53218.6 n + 133972 \biggr), \qquad   
\label{eq:3} 
\end{eqnarray}
  
\begin{eqnarray}
\gamma^{-1} = 1 - {{n + 2} \over {n + 8}}~g 
+ {g^2 \over (n + 8)^2}~(n + 2)~3.375628955 
\qquad \qquad \qquad \qquad \qquad \qquad
\nonumber \\
- {g^3 \over (n + 8)^3} \biggl( 4.661884772 n^2 + 34.41848329 n 
+ 50.18942749 \biggr) \ \qquad \qquad \qquad \qquad 
\nonumber \\
+ {g^4 \over (n + 8)^4} \biggl( 0.3189930 n^3 + 71.703302 n^2 
+ 429.42449 n + 574.58772 \biggr) \ \ \qquad \qquad
\nonumber \\
- {g^5 \over (n + 8)^5} \biggl( - 0.11970 n^4 + 69.379 n^3 
+ 1482.76 n^2 + 6953.61 n + 8533.16 \biggr), \qquad
\label{eq.4}
\end{eqnarray}

\begin{eqnarray}
\eta = {g^2 \over (n + 8)^2}~(n + 2)~0.9170859698 
- {g^3 \over (n + 8)^2}~(n + 2)~0.05460897758 \qquad \quad
\nonumber \\
+ {g^4 \over (n + 8)^4} \biggl( - 0.09268446 n^3 + 4.0564105 n^2 
+ 29.251167 n + 41.535216 \biggr) \ \qquad
\nonumber \\
- {g^5 \over (n + 8)^5} \biggl( 0.07092 n^4 + 1.05240 n^3 
+ 57.7615 n^2 + 325.329 n + 426.896 \biggr); \qquad
\label{eq.5}
\end{eqnarray}
the refined expression for the five-loop contribution to 
$\gamma^{-1}$ is taken from [9]. Instead of the renormalized 
coupling constant $g_4$, a rescaled coupling  
\begin{equation}
g = {n + 8 \over {24 \pi}} g_4,
\label{eq:6}
\end{equation}
is used as an argument in above RG series. This variable is more 
convenient since it does not go to zero under $n \to {\infty}$ but 
approaches the finite value equal to unity.

To evaluate the Wilson fixed point location $g^*$ and numerical 
values of the critical exponents, the resummation procedure based 
on the Borel-Leroy transformation 
\begin{equation}
f(x) = \sum_{i = 0}^{\infty} c_i x^i = \int\limits_0^{\infty} 
e^{-t} t^b F(xt) dt, \ \ \ \ \   
\nonumber\\
F(y) = \sum_{i = 0}^{\infty} {c_i \over (i + b)!} y^i \ \ , 
\label{eq:7}
\end{equation}
is used. The analytical extension of the Borel transforms is 
performed by exploiting relevant Pad\'e approximants [L/M]. In 
particular, four subsequent diagonal and near-diagonal approximants 
$[1/1]$, $[2/1]$, $[2/2]$, and $[3/2]$ turn out to lead to numerical 
estimates for $g^*$ which rapidly converge, via damped oscillations, 
to the asymptotic values; this is cleary seen from Table 1. These 
asymptotic values, i. e. the final five-loop RG estimates for $g^*$ 
are presented in Table 2 for $0 \le n \le 8$ (to avoid confusions, 
let us note that models with $n \ge 2$ possessing no ordered phase 
are studied here only as polygons for testing the numerical power of 
the perturbative RG technique). As Table 2 demonstrates, the numbers 
obtained differ appreciably from numerical estimates for $g^*$ given 
by the lattice and Monte Carlo calculations [10-16]; such estimates 
are usually extracted from the data obtained for the linear ($\chi$) 
and non-linear ($\chi_4$) susceptibilities related to each another 
via $g_4$: 
\begin{equation}
\chi_4 = {\partial^3M \over{\partial H^3}} \Bigg\arrowvert_{H = 0} 
= - \chi^2 m^{-2} g_4, \qquad \quad 
\label{eq:8} \\
\end{equation}
Since the convergence of the numerical estimates for $g^*$ given 
by the resummed RG series is oscillatory, an 
account for higher-order (six-loop, seven-loop, etc.) terms in the 
expansion (3) will not avoid this discrepancy [9]. That is why we 
believe that it reflects the influence of the singular 
(non-analytical) contribution to the $\beta$-function. 

The critical exponents for the Ising model ($n = 1$) and for those 
with $n = 0$ and $n = -1$ are estimated by the Pad\'e-Borel summation 
of the five-loop expansions (4), (5) for $\gamma^{-1}$ and $\eta$. Both 
the five-loop RG (Table 1) and the lattice (Table 2) estimates for 
$g^*$ are used in the course of the critical exponent evaluation. 
To get an idea about an accuracy of the numerical results obtained 
the exponents are estimated using different Pad\'e approximants, 
under various values of the shift parameter $b$, etc. In particular, 
the exponent $\eta$ is estimates in two principally different ways: 
by direct summation of the series (5) and via the resummation of RG 
expansions for exponents
\begin{equation}
\eta^{(2)} = {1 \over \nu} + \eta - 2, 
\qquad \qquad \eta^{(4)} = {1 \over \nu} - 2,  
\label{eq:9}
\end{equation}
which possess a regular structure favouring the rapid convergence of 
the iteration procedure. The typical error bar thus found is about 
0.05. 

The results obtained are collected in Table 3. As is seen, for 
small exponent $\eta$ and in some other cases the differences between 
the five-loop RG estimates and known exact values of the critical 
exponents exceed the error bar mentioned. Moreover, in the five-loop 
approximation the correction-to-scaling exponent $\omega$ of the 2D 
Ising model is found to be close to the value 4/3 predicted by the 
conformal theory [17] and to the estimate $1.35 \pm 0.25$ extracted 
from the high-temperature expansions [18] but differs markedly from 
the exact value $\omega = 1$ [19] and contradicts to the  
conjecture $\omega = 2$ [20]. This may be considered as an argument 
in favour of the conclusion that non-analytical contributions are 
visible in two dimensions.  

The field theory enables us also to find the higher-order,  
sextic coupling constant entering the free energy expansion in powers 
of the magnetization $M$. For 2D Ising model 
\begin{equation}
F(M) - F(0) = m^2 \Biggl[{1 \over 2}{M^2 \over Z} + 
g_4 \Biggl({M^2 \over Z} \Biggr)^2 +
\sum_{k=3}^{ \infty} {g_{2k} \Biggl({M^2 \over Z} \Biggr)^k} \Biggr],
\label{eq:10}
\end{equation}
where $m$ is a renormalized mass, $Z$ being a field renormalization 
constant. In the critical region, where fluctuations are so strong 
that they completely screen out the initial (bare) interaction, the 
behaviour of the system becomes universal and dimensionless 
effective couplings $g_{2k}$ approach their asymptotic limits 
$g_{2k}^*$.

In order to estimate $g_6^*$ we calculate RG expansion for $g_6$ 
and then apply Pade-Borel-Leroy resummation technique to get proper 
numerical results. As is well known, accurate enough numerical 
estimates may be extracted only from sufficiently long RG series. 
We have obtained the expression for $g_6$ in the four-loop 
approximation [21] which turned out to provide fair numerical 
estimates for the quantity of interest.

The method of calculating the RG series we used in [21] is 
straightforward. Since in two dimensions higher-order bare 
couplings are irrelevant in RG sense, renormalized
perturbative series for $g_6$ can be obtained from conventional
Feynman graph expansion of this quantity in terms of the only 
bare coupling constant - quartic coupling $\lambda$. In its turn, 
$\lambda$ may be expressed perturbatively as a function of 
renormalized dimensionless quartic coupling constant $g_4$. 
Substituting corresponding power series for $\lambda$ into original 
expansion we can obtain the RG series for $g_6$. As was shown 
[22,23], the one-, two-, three- and four-loop contributions are 
formed by 1, 3, 16, and 94 one-particle irreducible Feynman graphs, 
respectively. Their calculation along with the renormalization 
procedure just described gives:
\begin{equation}
g_6 = {36 \over \pi}\ {g_4^3}\ {\Bigl( 1 - 3.2234882\ g_4 + 
14.957539\ g_4^2 - 85.7810\ g_4^3 \Bigr)}.
\label{eq:11}
\end{equation}
This series may be used for estimation of the universal number 
$g_6^*$.

With the four-loop expansion in hand, we can construct three 
different Pad\'e approximants: [2/1], [1/2], and [0/3]. To 
obtain proper approximation schemes, however, only diagonal [L/L] 
and near-diagonal Pade approximants should be employed. That's why 
further we limit ourselves with approximants 
[2/1] and [1/2]. Moreover, the diagonal Pad\'e approximant [1/1] is  
also dealt with although this corresponds, in fact, to the usage 
of the lower-order, three-loop RG approximation. 

Since the Taylor expansion for the free energy contains as 
coefficients the ratios $R_{2k} = g_{2k}/g_4^{k-1}$ rather than 
the renormalized coupling constants themselves:
\begin{equation}
F(z) - F(0) = {m^2 \over g_4} \Biggl({z^2 \over 2} + z^4 + R_6 z^6 
+ R_8 z^8 + ... \Biggr), \qquad \qquad z^2 = {g_4 M^2 \over Z},  
\label{eq:12}
\end{equation}
we work with the RG series for $R_6$. It is resummed in three 
different ways based on the Borel-Leroy transformation and the 
Pade approximants just mentioned. The Borel-Leroy integral is 
evaluated as a function of the parameter $b$ under $g_4 = g_4^*$. 
For the fixed point coordinate the value $g_4^* = 0.6124$ [24] 
is adopted which is believed to be the most accurate estimate 
for $g_4^*$ available nowadays. The optimal value 
of $b$ providing the fastest convergence of the iteration scheme is 
then determined. It is deduced from the condition that the 
Pade approximants employed should give, for $b = b_{opt}$, 
the values of $R_6^*$ which are as close as possible to each other. 
Finally, the average over three estimates for $R_6^*$ is found and 
claimed to be a numerical value of this universal ratio. 

The results of our calculations are presented in Table 4. 
As one can see, for $b = 1.24$ all three working approximants 
lead to practically identical values of $R_6^*$. Hence,
we conclude that for 2D Ising model at criticality   
\begin{equation}
R_6^* = 2.94, ~~~~~~~g_6^* = 1.10. \ \
\label{eq:13}
\end{equation}
How close to their exact counterparts may these numbers be?  
To clear up this point let us discuss the sensitivity of numerical 
estimates given by RG expansion (11) to the type of 
resummation. The content of Table 4 implies that, among others, 
the results given by Pad\'e approximant $[2/1]$ turn out to 
be most strongly dependent on the parameter $b$. This situation 
resembles that for 3D $O(n)$-symmetric model where Pad\'e  
approximants of $[L-1/1]$ type for $\beta$-function and critical 
exponents lead to numerical estimates demonstrating 
appreciable variation with $b$ while for diagonal and near-diagonal 
approximants the dependence of the results on the shift parameter 
is practically absent [1,3,25]. In our case, Pad\'e approximants 
$[1/1]$ and $[1/2]$ may be referred to as generating such 
"stable" approximations for $g_6^*$. Since for $b$ varying from 0 
to 15 (i.e., for any reasonable $b$) the magnitude of $g_6^*$ 
averaged over these two approximations remains within the segment 
(1.044, 1.142) it is hardly believed that the values (13) can 
differ from the exact ones by more than 5$\%$.   

Another way to judge how accurate our numerical results 
are is based on the comparison of the values of $g_6^*$ given by 
four subsequent RG approximations available. While within the 
one-loop order we get $g_6^* = 2.633$ which is obviously very bad 
estimate, taking into account of higher-order RG contributions 
to $g_6$ improves the situation markedly. Indeed,  two-, three-, 
and four-loop RG series when resummed by means of the Pad\'e-Borel 
technique with use of "most stable" approximants $[0/1]$, $[1/1]$, 
and $[1/2]$ yield for $g_6^*$ the values 0.981, 1.129, and 1.051, 
respectively. Since this set of numbers demonstrates an oscillatory 
convergence one may expect that the exact value of renormalized 
sextic coupling constant lies between the higher-order -- three-loop 
and four-loop -- estimates. It means that the deviation of numbers 
(13) from the exact values would not exceed 0.05.

It is instructive to compare our estimates with those obtained by 
other methods. S.-Y. Zinn, S.-N. Lai, and M. E. Fisher
analyzing high temperature series for various 2D Ising lattices found
that $R_6^* = 2.943 \pm 0.007$ [26]; almost identical value 
was obtained in [27]. Our result for $R_6^*$ is seen to be in 
a brilliant agreement with this number. Of course, practical 
coincidence of the lattice and four-loop RG estimates is occasional 
and can not be considered as a manifestation of extremely high 
accuracy of the methods discussed. The closeness of these estimates 
to each another, however, unambiguously demonstrates high power of 
both approaches. Moreover, such a closeness shed a light on the role 
of a singular contribution to $g_6$ which can not be found 
perturbatively: this contribution is seen to be numerically small.   

It is interesting also to address the results given by another 
field-theoretical approach -- the $\epsilon$ expansion. For the Ising 
systems three terms in the $\epsilon$ expansion for 
$R_6$ are known [28]:
\begin{equation}
R_6^* = 2 \epsilon \Biggl(1 - {10 \over {27}}\ \epsilon + 
0.63795\ {\epsilon}^2 \Biggr). 
\label{eq:14}
\end{equation}   
Let us apply a simple Pad\'e-Borel procedure to this series 
as a whole and to the series in brackets and then put $\epsilon = 2$. 
We find $R_6^* = 3.19$ and $R_6^* = 3.12$ respectively, i.e. the numbers 
which differ from our estimate by less that 9$\%$. Keeping in mind lack 
of a small parameter these values of $R_6^*$ may be referred to as 
consistent. Proper account for higher-order terms in the 
$\epsilon$ expansion for $R_6$ should make corresponding numerical 
estimates closer to those extracted from 2D RG and high-temperature 
series. Very good agreement between the first number (13) and 
the estimate $R_6^* = 2.95 \pm 0.03$ [27] obtained by matching of the 
$\epsilon$ expansion with the exact results known for $D = 1$ and 
$D = 0$ may be considered as an argument in favor of this belief. 
One can find more details in recent comprehensive review [29]. 

Along with the RG calculations at physical dimension and the 
$\epsilon$ expansion, some other field-theoretical approach may be 
employed to estimate the critical parameters of 2D Ising model. 
We mean the method of the pseudo-$\epsilon$ expansion (see Ref. 19 
in [2]). Pseudo-$\epsilon$ expansions for the Wilson fixed point 
coordinate $g^*$ and critical exponents can be easily derived from 
the RG series (3)-(5) using standart technique. They are as follows 
[30]: 
\begin{equation}
g^* = \tau + 0.716173621 \tau^2 + 0.095042867 \tau^3  
+ 0.086080396 \tau^4 - 0.204139 \tau^5 ,  
\label{eq:15}
\end{equation} 

\begin{equation}
\gamma^{-1} = 1 - {\frac{1}{3}} \tau - 0.113701246 \tau^2  
+ 0.024940678 \tau^3 - 0.039896059 \tau^4 + 0.0645212 \tau^5 , 
\label{eq:16}
\end{equation}

\begin{equation}
\eta = 0.033966147 \tau ^2 + 0.046628762 \tau^3 
+ 0.030925471 \tau^4 + 0.0256843 \tau^5 . 
\label{eq:17}
\end{equation}
Note that the higher-order terms in series (15) and (16) have small  
numerical coefficients and irregular signs. Smallness of these 
coefficients enables one to obtain accurate enough estimates for 
$g^*$ and critical exponent $\gamma$ without addressing the Borel- 
transformation-based resummation methods. 

To demonstrate this, conventional Pade triangles originating from 
(15) and (16) under $\tau = 1$ are presented here (Tables 5 and 6).  
Since diagonal and near-diagonal Pad\'e approximants are known to 
exhibite the best approximating properties, the numbers 1.751 and 
1.837 from Table 5 should be referred to as most reliable estimates 
for $g^*$. Averaging over them, we obtain $g^* = 1.794$ which differs 
from the exact value $g^* = 1.75436$ [24] by 2\%. As seen from 
Table 5, it is the five-loop approximation that provides so good 
numerical result; almost all lower-order approximations suffers 
from dangerous poles resulting in strongly scattered estimates. 
The same is true for the susceptibility exponent. Indeed, the 
numbers given by the main working approximants [2/3] and [3/2], as 
well as by approximant [4/1], are almost coincide with each other 
and are close to the exact value $\gamma = 1.75$. In contrast, 
approximants [2/2] and [1/3], corresponding to the four-loop order, 
have dangerous poles which considerably affect the results. 

Unfortunately, the pseudo-$\epsilon$ expansion technique turns out 
to be much less powerful when applyed to estimate "small" critical 
exponent $\eta$. Both the direct summation of the expansion (17) 
and Pad\'e resummation of the series for "big" exponents $\gamma$ 
and $\nu$ lead to the numbers differing by 0.1 and even more from 
the exact value $\eta = 0.25$ [30]. To the contrary, the 
pseudo-$\epsilon$ expansion for the ratio $R_6 = g_6/{g_4^2}$ 
\begin{equation}
R_6 = 4 \tau (1 - 0.409036 \tau + 0.305883 \tau^2 
- 0.437676 \tau^3) \
\label{eq:18}
\end{equation}
demonstrates good Pad\'e summability. It is clearly seen from 
Table 7 [30]. Averaging over two working approximants [2/2] and 
[3/1] gives the number $R_6 = 2.90$ which is close to earlier 
estimates $R_6 = 2.94$ [21], $R_6 = 2.95$ [27], $R_6 = 2.943$ [26], 
and to high-precision values $R_6 = 2.94294$ [24], $R_6 = 2.94238$ 
[29,31]. Usage of more advanced, Pad\'e-Borel resummation technique 
shifts the pseudo-$\epsilon$ expansion estimate to $R_6 = 2.94$ 
[30] making it practically equal to just mentioned numbers. 

The area where 2D $\lambda \phi^4$ field theory can be successfully  
applyed is not limited by Ising-like and 0(n)-symmetric systems. 
The RG analysis of 2D cubic, MN, chiral, and weakly disordered models 
proofs to be rather effective provided the higher-order -- four- and 
five-loop -- approximations are used [9,32-34]. In particular, 
many-loop RG calculations reproduce with high accuracy the exact 
results known for 2D anisotropic systems with n-vector order 
parameters. Detailed description of the situation may be found in 
[9,33,34].        

I thank P. Calabrese, D. V. Pakhnin, P. Parruccini, and E. V. Orlov 
for fruitful collaboration. I am also grateful to B. N. Shalaev for 
numerous valuable discussions of the critical thermodynamics of 2D 
systems. This work was supported by the Russian Foundation for 
Basic Research under Grant No. 04-02-16189.

\newpage
\begin{table}
\caption{The Wilson fixed point coordinate for models with $n = 1$, 
$n = 0$ and $n = -1$ in four subsequent RG approximations and the 
final five-loop estimates for $g^*(n)$.}
\begin{tabular}{|c|c|c|c|c|c|}
\hline
~~~$n$~~~& [1/1]~~~& [2/1]~~~& [2/2]~~~& [3/2]~~~& $g^*$, 5-loop~~\\
\hline
\hline
1  & 2.4246~~~~& 1.7508~~~~& 1.8453~~~~& 1.8286~~~~& 1.837 $\pm$ 
0.03~~~~\\
\hline
0  & 2.5431~~~~& 1.7587~~~~& 1.8743~~~~& 1.8402~~~~& 1.86 $\pm$ 
0.04~~~~\\
\hline
-1 & 2.6178~~~~& 1.7353~~~~& 1.8758~~~~& 1.8278~~~~& 1.85 $\pm$ 
0.05~~~~\\
\hline
\end{tabular}
\label{table1}
\end{table}

\begin{table}
\caption{The Wilson fixed point coordinate $g^*$ and critical exponent  
$\omega$ for $0 \le n \le 8$ obtained in the five-loop RG 
approximation. The values of $g^*$ extracted from high-temperature (HT) 
[11,13] and strong coupling (SC) [12] expansions, found by Monte Carlo 
simulations (MC) [14,15], obtained by the constrained resummation of 
the $\epsilon$-expansion for $g^*$ ($\epsilon$-exp.) [13], and given 
by corresponding $1/n$-expansion ($1/n$-exp.) [13] are also presented 
for comparison.}
\begin{tabular}{|c|c|c|c|c|c|c|}
\hline
$n$ & 0 & 1 & 2 & 3 & 4 & 8 \\
\hline
\multicolumn{7}{|c|}{$g^*$} \\
\hline
RG, 5-loop & 1.86(4) & 1.837(30) & 1.80(3) & 1.75(2) 
& 1.70(2) & 1.52(1) \\
&  &  &  &  & ($b=1$) & ($b=1$) \\
\hline
HT & 1.679(3) & 1.754(1) & 1.81(1) & 1.724(9) 
& 1.655(16) &  \\
\hline
MC &  & 1.71(12) & 1.76(3) & 1.73(3) &  &  \\  
\hline
SC & 1.673(8) & 1.746(8) & 1.81(2) & 1.73(4) &  &  \\
\hline
$\epsilon$-exp. & 1.69(7) & 1.75(5) & 1.79(3) & 1.72(2) 
& 1.64(2) & 1.45(2) \\
\hline
1/n-exp. &  &  &  & 1.758 & 1.698 & 1.479 \\
\hline
\hline
\multicolumn{7}{|c|}{$\omega$} \\
\hline
RG, 5-loop & 1.31(3) & 1.31(3) & 1.32(3) & 1.33(2) 
& 1.37(3) & 1.50(2) \\
\hline
\end{tabular}
\label{table2}
\end{table}

\begin{table}
\caption{Critical exponents for $n = 1$, $n = 0$, and $n = -1$ 
obtained via the Pad\'e-Borel summation of the five-loop RG 
expansions for $\gamma^{-1}$ and $\eta$. The known exact values 
of these exponents are presented for comparison.}
\begin{tabular}{|c|c|c|c|c|c|c|}
\hline
~~~$n$~~~&  & $g^*$ & $\gamma$ & $\eta$ & $\nu$ & $\alpha$ \\
\hline
\hline
1  & RG    & 1.837        & 1.79 & 0.146 & 0.96~~~& 0.07~~~ \\
   &       & 1.754 (HT)~~~& 1.74 & 0.131 & 0.93~~~& 0.14~~~ \\
\hline   
   & exact~~&              &  7/4   &  1/4   &  1     &  0     \\   
   &       &              & (1.75) & (0.25) &        &         \\
\hline   
\hline
0  & RG    & 1.86         & 1.45 & 0.128 & 0.77 & 0.45 \\
   &       & 1.679 (HT)~~~& 1.40 & 0.101 & 0.74 & 0.52 \\
\hline
   & exact~~& & 43/32        & 5/24         & 3/4~~  & 1/2~~ \\
   &       & & (1.34375)~~~~& (0.20833)~~~~& (0.75) & (0.5) \\
\hline
\hline
-1 & RG    & 1.85         & 1.18 & 0.082 & 0.62 & 0.76 \\
   &       & 1.473 (SC)~~~& 1.15 & 0.049 & 0.59 & 0.82 \\
\hline   
   & exact~~& & 37/32        &  3/20  &   5/8   &  3/4   \\
   &       & & (1.15625)~~~~& (0.15) & (0.625) & (0.75)  \\
\hline
\end{tabular}
\label{table3}
\end{table}

\begin{table}
\caption{The values of $R_6^*$ obtained by means of the 
Pad\'e-Borel-Leroy technique for various $b$ within 
three-loop (approximant $[1/1]$) and four-loop (approximants 
$[1/2]$ and $[2/1]$) RG approximations. The estimate for $b = 1$ 
in the middle line is absent because corresponding Pad\'e 
approximant turnes out to be spoilt by a positive axis pole.}
\begin{tabular}{ccccccccccc}
$b$ & 0 & 1 & 1.24 & 2 & 3 & 4 & 5 & 7 & 10 & 15 \\
\tableline
$[1/1]$ & 2.741 & 2.908 & 2.937 & 3.009 & 3.077 & 3.125 & 3.161 
& 3.212 & 3.258 & 3.301 \\
\tableline
$[1/2]$ & 2.827 &  -  & 2.936 & 2.877 & 2.853 & 2.838 & 2.828 
& 2.814 & 2.800 & 2.787 \\
\tableline
$[2/1]$ & 3.270 & 2.988 & 2.936 & 2.800 & 2.667 & 2.568 & 2.491 
& 2.380 & 2.273 & 2.171 \\
\end{tabular}
\label{table4}
\end{table}

\begin{table}
\caption{The Wilson fixed point location $g^*$, extracted from 
pseudo-$\epsilon$ expansion (15) by means of constructing Pad\'e 
approximants [L/M]. Coordinates of "dangerous" poles of Pad\'e 
approximants, i. e. those lying on the real positive semiaxis 
are indicated as subscripts.}
\begin{tabular}{|c|c|c|c|c|c|}
M~~~L & 1~~~~~~~~& 2~~~~~~~~& 3~~~~~~~~& 4~~~~~~~~
& 5~~~~~~~~\\
\hline
0 & 1.000~~~~~~~~~& 1.716~~~~~~~~~& 1.811~~~~~~~~~& 1.897~~~~~~~~~
& 1.693~~~~~~~~~ \\
\hline
1 & $3.523_{1.4}$~~~~~~& $1.826_{7.5}$~~~~~~& $2.724_{1.1}$~~~~~
& 1.837~~~~~~~~~~&  \\
\hline
2 & 1.425~~~~~~~~~& $1.918_{3.0}$~~~~~~& $1.850_{6.1}$~~~~~~&    &    \\
\hline
3 & $2.601_{1.4}$~~~~~~& 1.751~~~~~~~~~&       &       &       \\
\hline
4 & 1.194~~~~~~~~~&             &             &       &       \\
\end{tabular}
\label{table5}
\end{table}

\begin{table}
\caption{Numerical values of the critical exponent $\gamma$ obtained  
by Pad\'e summation of series (16) for $\gamma^{-1}$.} 
\begin{tabular}{|c|c|c|c|c|c|c|}
M~~~L & 0~~~~~~& 1~~~~~~& 2~~~~~~& 3~~~~~~& 4~~~~~~& 5~~~~~~\\
\hline
0 & 1.000~~~~~~~& 1.500~~~~~~~& 1.808~~~~~~~& 1.730~~~~~~~& 1.859~~~~~~~
& 1.660~~~~~~~\\
\hline
1 & 1.333~~~~~~~& $2.024_{2.9}$~~~~& 1.744~~~~~~~& 1.778~~~~~~~
& 1.777~~~~~~~&        \\
\hline
2 & 1.558~~~~~~~& 1.702~~~~~~~& $1.800_{5.2}$~~~~& 1.777~~~~~~~&  &  \\
\hline
3 & 1.646~~~~~~~& $6.871_{1.1}$~~~~& 1.772~~~~~~~&      &     &   \\
\hline
4 & 1.732~~~~~~~& 1.718~~~~~~~&        &       &       &        \\
\hline
5 & $1.714_{6.1}$~~~~&      &        &       &       &        \\
\end{tabular}
\label{table6}
\end{table}

\begin{table}
\caption{Pad\'e triangle for the universal ratio $R_6$ given by 
pseudo-$\epsilon$ expansion (18).}
\begin{tabular}{|c|c|c|c|c|}
M~~~L & 1~~~~~~~~~~~~& 2~~~~~~~~~~~~& 3~~~~~~~~~~~~& 4~~~~~~~~~~~~ \\
\hline
0 & 4.000~~~~~~~~~~~~~& 2.364~~~~~~~~~~~~& 3.587~~~~~~~~~~~~& 1.837~~~~~~~~~~~~ \\
\hline
1 & 2.839~~~~~~~~~~~~~& 3.064~~~~~~~~~~~~& 2.867~~~~~~~~~~~~&       \\
\hline
2 & $3.148_{4.5}$~~~~~~~~~~& 2.940~~~~~~~~~~~~&      &       \\
\hline
3 & 2.621~~~~~~~~~~~~~&       &        &       \\
\end{tabular}
\label{table7}
\end{table}

\end{document}